\documentclass{PoS}

\usepackage{amsmath}

\title{Improved analysis of the scalar and vector form factors of kaon
semileptonic decays with $N_f = 2$ twisted-mass fermions}

\ShortTitle{Improved analysis of the scalar and vector form factors of
kaon semileptonic decays}

\author{ETM Collaboration}

\author{ V.~Lubicz$^{ab}$, F.~Mescia $^{c}$, \speaker{L.~Orifici} $^{ab}$,
              S.~Simula $^{b}$ and C.~Tarantino $^{ab}$\\%
             \llap{$^{a}$} Dipartimento di Fisica, Universit\`a Roma Tre,
                      Via della Vasca Navale 84, 00146 Roma, Italy\\
            \llap{$^{b}$} INFN - Sezione Roma Tre,
                     Via della Vasca Navale 84, 00146 Roma, Italy\\
            \llap{$^{c}$} Dep. ECM and ICC, Universitat de Barcelona, 
                     Diagonal 647, 08028 Barcelona, Spain\\
}

\abstract{We investigate the vector and scalar form factors relevant for $K_{\ell 3}$ semileptonic 
decays using maximally twisted-mass fermions with two flavors of dynamical quarks ($N_f = 2$).
The simulations cover pion masses as light as 260 MeV and four values of the lattice spacing, 
ranging from $\sim$~0.05 up to $\sim$ 0.1 fm, which allow to compute directly, for the first time, 
the continuum limit for the vector form factor at zero-momentum transfer, $f_+(0)$.
The preliminary result is $f_+(0) = 0.9544~(68_{stat})$, 
where the error is statistical only.
We also extrapolate both form factors to the physical point and study their momentum dependence.
Our results are in good agreement with those obtained from a dispersion analyses of the 
experimental data. Together with the form factors, we analyze the ratio of the leptonic decay constants $f_K / f_\pi$, by imposing the constraint coming from the Callan-Treiman theorem, obtaining at the physical point  $f_K / f_\pi = 1.190 ~ (8_{stat})$. 
Combining our results for $f_+(0)$ and $f_K / f_\pi$ with the experimental measurements of the leptonic and semilpetonic decay rates, and using the determination of $|V_{ud}|$ from nuclear beta decays, we determine the values of the Cabibbo angle $|V_{us}|$ from both $K_{\ell 3}$ and $K_{\ell 2}$ decays, obtaining $|V_{us}|^{K_{\ell 3}} = 0.2266~(17)$ 
and $|V_{us}|^{K_{\ell 2}} = 0.2258~(16)$. 
}

\FullConference{The XXVIII International Symposium on Lattice Field Theory, Lattice2010\\
		June 14-19, 2010\\
		Villasimius, Italy}

\begin{document}

\section{Introduction}

Weak hadron decays are very interesting processes because measuring the decay widths for such processes allows us to extract some of the fundamental parameters of the Standard Model, namely the entries of the Cabibbo-Kobayashi-Maskawa (CKM) quark-mixing matrix \cite{CKM}. The needed theoretical inputs from non-perturbative QCD are the form factors and decay constants which parametrize the hadronic matrix elements relevant for each decay.

In the case of the $K \to \pi \ell \nu_{\ell}$ semileptonic decay, the matrix element of the weak vector current can be written in terms of two form factors, the vector, $f_{+}(q^{2})$, and the scalar, $f_{0}(q^{2})$, form factors:
 \begin{eqnarray}
      \langle \pi(p_{\pi}) | V^{\mu} | K(p_{K}) \rangle = (p_{\pi} + p_{K} - \Delta)^{\mu} ~ f_{+}(q^2) 
     + \Delta^{\mu} ~ f_{0}(q^2) ~ ,
     \label{eq:Kpi_ME}
 \end{eqnarray}
where $\Delta^\mu \equiv q^\mu ~ (M_K^2 - M_\pi^2) / q^2$ and  $q^\mu \equiv (p_K - p_\pi)^\mu$ is the 4-momentum transfer. This decay is relevant for the determination of the \textit{CKM} matrix element 
$|V_{us}| \simeq \sin \theta$, where $\theta$ is the Cabibbo angle.

In this contribution we present a lattice study of the vector and the scalar form factors performed by using the gauge configurations generated by the European Twisted Mass Collaboration (ETMC) with $N_f = 2$ maximally twisted-mass fermions. We present our preliminary results coming from two different strategies. The first one is a direct update of the analysis of Ref.~\cite{ETMC_Kl3}, which was based on simulations at two values of the lattice spacing ($a \simeq 0.086$ and $a \simeq 0.068$ fm). Here we employ the results of simulations performed at four values of the lattice spacing, ranging from $a \sim $ 0.05 up to $\sim$ 0.1 fm. This allows us to compute for the first time, in a well controlled way, the continuum limit for the vector form factor at zero-momentum transfer, $f_+(0)$. The second strategy is a multi-combined fit of the $q^2$, $M_\pi$ and \textit{a} dependencies of both the vector and the scalar form factors. We also analyze, together with the form factors, the ratio of the leptonic decay constants $f_K / f_\pi$, by imposing the constraint coming from the Callan-Treiman (CT) theorem \cite{CT}. In this way we determine the full momentum dependence of the form factors at the physical point, finding results which agree nicely with those obtained from a recent dispersion analyses of the experimental data \cite{FLAVIANET}. Our preliminary results for $f_+(0)$ and $f_K / f_\pi$ are
  \begin{eqnarray}
      f_+(0) = 0.9544 ~ (68_{stat}) ~ , 
      \qquad f_K / f_\pi = 1.190 ~ (8_{stat}) ~ , 
      \label{eq:final_results}
  \end{eqnarray} 
where the quoted errors are statistical only. A detailed analysis of the systematic uncertainties, including the estimate of the quenching effect of the strange quark, is in progress and final results will be presented in a forthcoming publication.

Using the experimental information from $K_{\ell 3}$ and $K_{\ell 2}$ decays \cite{FLAVIANET},
together with the precise determination of $|V_{ud}|$ from superallowed nuclear beta decays, 
$|V_{ud}| = 0.97425 ~ (22)$ \cite{Vud}, we obtain the following determinations of the Cabibbo angle
  \begin{eqnarray}
      |V_{us}|^{K_{\ell 3}} = 0.2266 ~ (17), 
      \qquad |V_{us}|^{K_{\ell 2}} = 0.2258 ~ (16), 
  \end{eqnarray}
in good agreement among each other.

\section{First strategy}

We have performed the calculations of all the relevant 2-point and 3-point correlation functions using the ETMC gauge configurations with $N_f = 2$ dynamical twisted-mass quarks \cite{ETMC_scaling} generated at four values of $\beta$, namely the ensembles $A_2 - A_4$ at $\beta = 3.8$ ($a \simeq 0.101$ fm), $B_2 - B_7$ at $\beta = 3.9$ ($a \simeq 0.086$ fm), $C_2 - C_3$ at $\beta = 4.05$ ($a \simeq 0.068$ fm) and $D_2$ at $\beta = 4.2$ ($a \simeq 0.054$ fm). The pion mass $M_\pi$ ranges between $\simeq 260$ MeV and $\simeq 580$ MeV and the size $L$ of our lattices guarantees that  $M_\pi L$ is always larger than $4.0$ except for the ensemble $B_7$ ($M_\pi L \simeq 3.7$). For each pion mass and lattice spacing we have used several values of the (bare) strange quark mass $m_s$ to allow for a smooth, local interpolation of our results to the physical value of $m_s$ (see Ref.~\cite{ETMC_masses}).

Following Ref.~\cite{ETMC_Kl3}, the momentum dependence of the calculated form factors is fitted using either a pole or a quadratic behavior in order to determine the values of $f_+(0)$ at each simulated pion and kaon masses. Then our results for $f_+(0)$ can be smoothly interpolated (by quadratic splines) in terms of the kaon mass at a reference value $M_K^{ref}$, obtained by fixing the combination $(2M_K^2-M_\pi^2)$ at its physical value.

\begin{figure}[!hbt]
  \centerline{\includegraphics[width=0.7\textwidth]{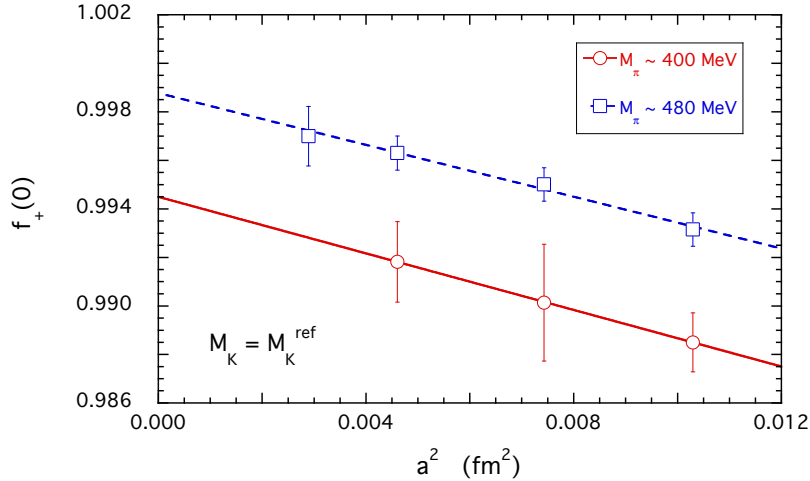}}
  \caption{\it Results for $f_+(0)$ versus $a^2$ for $M_\pi \simeq 400$ and $480$ MeV at $M_K = 
  M_K^{ref}$. The dashed and solid lines represent linear fits in $a^2$.}
  \label{fig:f0_a2}
\end{figure}

Discretization effects are found to scale linearly with $a^2$, as shown in Fig.~\ref{fig:f0_a2}, in agreement with the expectation of (automatic) ${\cal{O}}(a)$-improvement of the maximally twisted-mass formulation \cite{improvement}.

\begin{figure}[!hbt]
  \centerline{\includegraphics[width=0.7\textwidth]{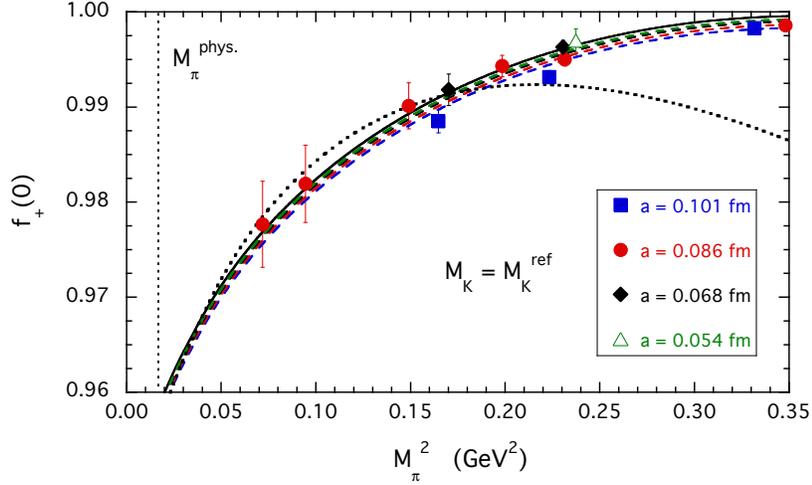}} 
  \caption{\it Results for $f_+(0)$ versus $M_\pi^2$ at $M_K = M_K^{ref}$ for four values of the lattice spacing $a$. The dashed lines represent the SU(2) fit (\protect\ref{eq:SU2}) at each value of the lattice spacing, while the solid line is the same fit in the continuum limit. The dotted line is the fit (\protect\ref{eq:SU2}) at NLO (i.e., with $d_+ = 0$) applied to our data with $M_\pi \lesssim 400$ MeV and evaluated in the continuum limit. The vertical line corresponds to $M_\pi^{phys} = 135.0$ MeV.}
  \label{fig:f0}
\end{figure}

Thus, as in Ref.~\cite{ETMC_Kl3}, we perform the chiral extrapolation starting from the SU(2) ChPT 
prediction at next-to-leading order (NLO) \cite{SU2} and adding to it  both a NNLO term proportional to $M_\pi^4$, in order to analyze all our data up to $M_\pi \sim 580$ MeV, and a linear term in $a^2$ in order to correct for lattice artifacts, namely 
 \begin{eqnarray}
    f_+(0) = F_+ \left[ 1 - \frac{3}{4} \frac{M_\pi^2}{(4 \pi f)^2} \mbox{log}\left(\frac{M_\pi^2}{\mu^2}\right) 
                   + c_+ M_\pi^2 + d_+ M_\pi^4 + D a^2 \right] ~ .
    \label{eq:SU2}
 \end{eqnarray}
In Eq.~(\ref{eq:SU2}) $f$ is the pion decay constant in the SU(2) chiral limit, and $F_+$, $c_+$, $d_+$ are SU(2) low-energy constants (LEC's) functions of the strange quark mass $m_s$\footnote{The LEC $c_+$ depends also on the renormalization scale $\mu$, but the whole result (\ref{eq:SU2}) is independent on $\mu$.}.
The quality of the fit (\ref{eq:SU2}) applied to our data with $M_\pi \lesssim 580$ MeV is shown in Fig.~\ref{fig:f0}, where we also compare it with the continuum limit result of a pure NLO fit (i.e., with $d_+ = 0$) applied to our data with $M_\pi \lesssim 400$ MeV.

After correcting for the effect of quenching the strange quark, as discussed in Ref.~\cite{ETMC_Kl3} (namely, $f_+(0) - f_+^{PQ}(0) = -0.0058 ~ (28)$), our preliminary result for $f_+(0)$ at the physical point is
 \begin{eqnarray}
    f_+(0) = 0.9542 ~ (60_{stat}) ~, 
    \label{eq:f0_1}
 \end{eqnarray}
where the quoted error is statistical only. A final result including the estimate of the systematic uncertainties due to residual discretization effects, momentum and quark mass extrapolation, and the quenching effect of the strange quark, will be presented in a forthcoming publication.
The preliminary value obtained for $f_+(0)$ agrees very well with our previous result of Ref.~\cite{ETMC_Kl3}, as well as with both the Leutwyler-Roos result \cite{LR} and previous lattice calculations performed with $N_f = 0$ \cite{SPQCDR}, $N_f = 2$ \cite{JLQCD_2005,RBC_2006,QCDSF_2007} and $N_f = 2 + 1$ \cite{RBC_2008} dynamical flavours.

\section{Second strategy}

The second analysis consists of performing a multi-combined fit of the $q^2$, $M_\pi$ and $a$ 
dependencies of the form factors in order to predict, at the physical point, both the vector and the scalar form factors not only at $q^2 = 0$, but also in the entire $q^2$-region spanned by the experiments, i.e.~from $q^2 = 0$ to $q^2 =q_{max}^2 \equiv (M_K - M_\pi)^2$.

We consider, for the vector and the scalar form factors, the following functional forms
\begin{eqnarray}
   \label{eq:ff_SU2}
   f_+(s) & = & F_+(s) \left\{ 1 + C_+(s) x + \frac{M_K^2}{(4\pi f)^{2}} \left[ -\frac{3}{4} x \log{x} - 
   x T_1^+(s) - T_2^+(s) \right] \right\} ~ , \nonumber \\
   f_0(s) & = & F_0(s) \left\{ 1 + C_0(s) x + \frac{M_K^2}{(4\pi f)^2} \left[ -\frac{3}{4} x \log{x} + 
   x T_1^0(s) - T_2^0(s) \right] \right\} ~ ,
 \end{eqnarray}
where $x  = M_\pi^2 / M_K^2$ and $s  = q^2 / M_K^2$. The terms in the square brackets in Eq.~(\ref{eq:ff_SU2}) are derived from the NLO SU(3) ChPT predictions \cite{SU3} for the kaon and pion loop contributions to the form factors expanded in powers of $x$, keeping only the $\mathcal{O}(x)$, $\mathcal{O}(x \log x)$ and $\mathcal{O}(\log(1 - s))$terms. The functions $T_{1,2}^{0,+}(s)$ are then given by
 \begin{eqnarray}
   T_1^+(s) & = & \left[ (1 - s) \log{(1 - s)} + s(1 - s/2) \right] 3(1 + s) / 4s^2 ~ , \nonumber \\ 
   T_2^+(s) & = &  \left[ (1 - s) \log{(1 - s)} + s(1 - s/2) \right] (1 - s)^2 / 4s^2 ~ , \nonumber \\
   T_1^0(s) & = & \left[  \log{(1 - s)} + s(1 + s/2) \right] (9 + 7s^2) / 4s^2 ~ , \nonumber \\
   T_2^0(s) & = & \left[ (1 - s) \log{(1 - s)} + s(1 - s/2) \right] (1 - s) (3 + 5s) / 4s^2 ~ .
 \end{eqnarray}
It can be seen that the coefficients of the pion chiral log in Eq.~(\ref{eq:ff_SU2}) are in agreement with those predicted by SU(2) ChPT, both at $q^2 = 0$ and $q^2 = q_{max}^2$~\cite{SU2}. At $q^2 = 0$ the leading chiral log has the coefficient ($-3/4$), while close to $q^2 = q_{max}^2$, i.e. for $s \simeq (1 - \sqrt{x})^2$, the functions $T_{1,2}^0(s)$ also contribute to the chiral log leading, for $f_0(s)$, to an overall coefficient equal to ($-11/4$).

The functions $F_{0,+}(s)$ and $C_{0,+}(s)$ in Eq.~(\ref{eq:ff_SU2}) are not predicted by SU(2) chiral symmetry. For $F_{0}(s)$, we include in our analysis the constraint coming from the CT theorem \cite{CT}, which states that the scalar form factor $f_0(q^2)$ at the (unphysical) CT point, defined as $q_{CT}^2 = M_K^2 - M_\pi^2$, differs from the ratio of the leptonic decay constants $f_K / f_\pi$ by terms which are proportional to the light quark masses, namely: $f_0(q^2 = M_K^2 - M_\pi^2) = f_K / f_\pi + \mathcal{O}(m_{u,d})$. Therefore, in the SU(2) chiral limit the scalar form factor $f_0(q^2)$ at $q_{CT}^2 = q_{max}^2 = M_K^2$ coincides with the ratio of the leptonic decay constants. Since the SU(2) chiral expansion of $f_K / f_\pi$ is given at NLO by
 \begin{eqnarray}
    \frac{f_K} {f_\pi} = \frac{f_K^0} {f} \left[1 + Bx + \frac{M_K^2}{(4 \pi f)^2} \frac{5}{4} x \log{x}  \right] ~ ,
    \label{eq:fKfPi}
 \end{eqnarray}
where $f_K^0$ is the SU(2) chiral limit of $f_K$, the CT theorem is equivalent to impose on $F_0(s)$ the constraint
 \begin{eqnarray}
     F_0(s = 1) = f_K^0 / f ~ .
 \end{eqnarray}
Inspired by the vector-meson dominance, we then adopt a pole behavior for $F_{0,+}(s)$ and a polynomial (quadratic) behavior for $C_{0,+}(s)$, namely
 \begin{eqnarray}
    F_{0,+}(s) = F / [1 - \lambda_{0,+} s], \qquad C_{0,+}(s) = C + C_{0,+}^{(1)} s + C_{0,+}^{(2)} s^2 ~ .
    \label{eq:LECs}
 \end{eqnarray}

Finally, we take into account discretization effects by adding linear terms in $a^2$ to the parameter $F$ and to the slopes $\lambda_{0,+}$ of Eq.~(\ref{eq:LECs}), as well as in the chiral expansion (\ref{eq:fKfPi}) of $f_K / f_\pi$. As in the case of the first strategy, when we include in the fit all our data up to $M_\pi \sim 580$ MeV, we also add in Eq.~(\ref{eq:ff_SU2}) a NNLO term of the form $D_{0,+}(s)~x^2$, by expanding $D_{0,+}(s) = D + D_{0,+}^{(1)} s + D_{0,+}^{(2)} s^2$. 

\begin{figure}[!hbt]
  \centerline{\includegraphics[width=0.9\textwidth]{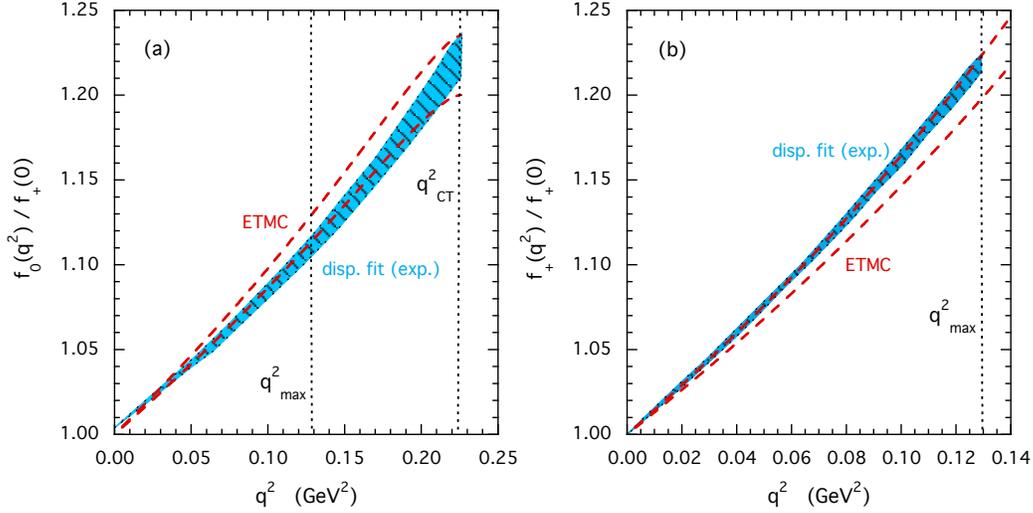}}
  \caption{\it The $q^2$-dependence of the scalar (a) and vector (b) form factors.  The dashed lines
   correspond to the region allowed at $1 \sigma$ by the fit of the lattice data according to 
   Eqs.~(\protect\ref{eq:ff_SU2}) and (\protect\ref{eq:fKfPi}). The shaded blue area is the $1\sigma$ region  
   obtained from a dispersive fit \cite{dispersive} of the KLOE, KTeV, NA48 (without muons branching ratios) 
   and ISTRA+ data performed in Ref.~\cite{FLAVIANET} . }
  \label{fig:f0_fp}
\end{figure}

Our analysis involves a total of $120$ data points with $16$ free parameters, and we obtain a good quality fit with $\chi^2 / d.o.f. \simeq 0.8$. The momentum dependence of the vector and scalar form factors extrapolated (for the first time) at the physical point is shown in Fig.~\ref{fig:f0_fp}. The lattice results are also compared in the plots with those obtained from a dispersive fit of the experimental data~\cite{FLAVIANET} from KLOE, KTeV, NA48 (without muons branching ratios) and ISTRA+, based on the parametrization of Ref.~\cite{dispersive}. It can be clearly seen that our results are in good agreement with the data, in the whole range of $q^2$ spanned by the experiments.

The preliminary results of this second strategy for the form factor at zero momentum transfer $f_+(0)$ and the ratio $f_K / f_\pi$ are
 \begin{eqnarray}
    f_+(0) = 0.9546 ~ (76_{stat}) ~, 
    \qquad f_K / f_\pi = 1.190 ~ (8_{stat}) ~, 
    \label{eq:f0_2}
 \end{eqnarray}
where, as in the case of the first strategy, the central value of $f_+(0)$ obtained from the fit has been shifted by $\delta f_+ = f_+(0) - f_+^{PQ}(0) = -0.0058$ in order to correct for the effect of quenching the strange quark.

\section{Results and conclusions}
Our preliminary best result for $f_+(0)$ is obtained by averaging the determinations (\ref{eq:f0_1}) and (\ref{eq:f0_2}) of the form factors obtained from the first and the second strategies respectively, leading to:
 \begin{eqnarray}
    f_+(0) = 0.9544 ~ (68_{stat}) ~ .    
    \label{eq:results1}
 \end{eqnarray}
Eqs~(\ref{eq:f0_2}) also provides our estimate of the ratio $f_K / f_\pi$ obtained by fitting the lattice data for the decay constants together with the semileptonic form factors and imposing the contraint coming from the CT theorem:
 \begin{eqnarray}
    f_K / f_\pi = 1.190 ~ (8_{stat}) ~ .
    \label{eq:results2}
 \end{eqnarray}
A careful analysis of the systematic uncertainties in the present calculation is still in progress. The results in Eqs.~(\ref{eq:results1}) and (\ref{eq:results2}) are in good agreement with the previous ETMC determination of the form factor, $f_+(0) = 0.9560(84)$~\cite{ETMC_Kl3}, based on simulations at only two values of the lattice spacing, and with the more extensive analysis of the meson decay constants presented in Ref~\cite{ETMC_fPS}, which quoted $f_K / f_\pi = 1.210(18)$. 

Combining our results (\ref{eq:results1}) and (\ref{eq:results2}) with the latest experimental averages \cite{FLAVIANET} $|V_{us}| \cdot f_{+}(0) = 0.2163 ~ (5)$ from $K_{\ell 3}$ decays and $|V_{us} / V_{ud}| \cdot f_K / f_\pi = 0.2758 ~ (5)$ from $K_{\ell 2}$ decays, and with the determination $|V_{ud}| = 0.97425 ~ (22)$ from nuclear beta decays~\cite{Vud}, we obtain for the Cabibbo angle the values
 \begin{eqnarray}
   |V_{us}|^{K_{\ell 3}} = 0.2266 ~ (17) ~ , \qquad |V_{us}|^{K_{\ell 2}} = 0.2258 ~ (16) ~ .
   \label{eq:Vus}
 \end{eqnarray}
 
The average of the $K_{\ell 3}$ and $K_{\ell 2}$ results in Eq.~(\ref{eq:Vus}) can be further combined with $V_{ud} = 0.97425 ~ (22)$ and $V_{ub} = 0.00376~(20)$~\cite{utfit} to test the unitarity of the first row of the CKM matrix, for which we find
  \begin{eqnarray}
     |V_{ud}|^2 + |V_{us}|^2 + |V_{ub}|^2 = 1.0003 ~ (8) ~ .
 \end{eqnarray}

\end{document}